\documentclass[11pt]{article}
\pdfoutput=1
 \usepackage{mciteplus}
 \usepackage{tikz}
 \usepackage{color}
 \definecolor{darkblue}{rgb}{0.1,0.1,.7}
 \usepackage[colorlinks, linkcolor=darkblue, citecolor=darkblue, urlcolor=darkblue, linktocpage,hyperfootnotes=false]{hyperref} 
\usepackage{epsfig}
\usepackage{graphicx}
\usepackage{cite}
\usepackage{amsfonts}
\usepackage{amssymb}
\usepackage{bm}
\usepackage{latexsym}
\usepackage{amsmath}
\numberwithin{equation}{section}
\def\bq{\begin{quote}}
\def\eq{\end{quote}}
 at 10truept

\newcommand{\calo}{{\cal O}}
\newcommand{\calh}{{\cal H}}
\newcommand{\cala}{{\cal A}}

\newcommand{\beq}{\begin{equation}}
\newcommand{\eeq}{\end{equation}}
\newcommand{\beqa}{\begin{eqnarray}}
\newcommand{\eeqa}{\end{eqnarray}}
\newcommand{\bea}{\begin{eqnarray}}
\newcommand{\eea}{\end{eqnarray}}

\def\roughly#1{\raise.3ex\hbox{$#1$\kern-.75em\lower1ex\hbox{$\sim$}}}

\begin{document}

\thispagestyle{empty}
\begin{titlepage}
 % \rightline{}
 \setcounter{page}{0}
  \bigskip

  \bigskip\bigskip

  \bigskip

\begin{center}
%\centerline
{\Large \bf {A ``black hole theorem," and its implications}}
    \bigskip
\bigskip
\end{center}

  \begin{center}

 \rm {Steven B. Giddings\footnote{\texttt{giddings@ucsb.edu}} }
  \bigskip \rm
\bigskip

{Department of Physics, University of California, Santa Barbara, CA 93106, USA}  \\
\rm

 % \bf {Write authors  }
  \bigskip \rm
\bigskip
 
\rm

\bigskip
\bigskip

% \vspace{2cm}
  \end{center}

\vspace{3cm}
  \begin{abstract}
  
A general formulation of the basic conflict of the information problem is given, encapsulated in a ``black hole theorem."
This is framed in a more general context than the usual one  of quantum field theory on a background, and is based on describing a black hole as a quantum subsystem of a larger system, including its environment.   This sharpens the limited set of possible consistent options; as with the Coleman-Mandula theorem, the most important point is probably the loophole in the ``theorem," and what this tells us about the fundamental structure of quantum gravity. 
This ``theorem" in particular involves the general question of how to define quantum subsystems in quantum gravity. If black holes do behave as quantum subsystems, at least to a good approximation, evolve unitarily, and do not leave remnants,
the ``theorem" implies the presence of interactions between a black hole and its environment that go beyond a description based on local quantum fields.   This provides further motivation for  and connects to previous work giving a principled parameterization of these interactions, and investigating their possible observational signatures via electromagnetic or gravitational wave observations of black holes.

 \medskip
  \noindent
  \end{abstract}
\bigskip \bigskip \bigskip 

  \end{titlepage}

\tableofcontents

\newpage

\section{Introduction}

Hawking's discovery\cite{Hawk} of black hole radiance led to the profound problem of black hole information.  This problem is plausibly  regarded as a key problem for quantum gravity, playing a role analogous to key problems such as the stability of the atom in the development of quantum mechanics.  Specifically, we would like to use this problem as a guide to the new principles of quantum gravity; an even more intriguing possibility is that its resolution will be associated with observational signatures.

Over time, there have been many proposed resolutions of this problem, though they fit into many fewer general categories.  And recently there has been a lot of focus on methods to calculate entropies\cite{AEMM,AMMZ,PSSY,AHMST,AHMSTrev}.  However, entropy curves are just {\it one} diagnostic of dynamics.  It is important to examine the problem -- and its resolution -- at a more basic level.

This short paper will suggest an organization of our understanding of the problem which focusses on the basic conflict,  connects to an information theoretic perspective, and connects to  what appear to be important questions regarding the basic structure of quantum gravity.  Aspects of the current perspective are certainly understood, either implicitly or explicitly, by many working in the field.  However, some things go beyond past discussion.

Specifically, firstly the present discussion will  place the problem in a more general context than that of evolution governed by local quantum field theory (LQFT) on a semiclassical background, in which it is typically formulated.  There are good reasons to expect that LQFT ultimately fails, and a very important question is what structure replaces it, and what of its features are retained.  If the world is fundamentally a quantum-mechanical system, a very general kinematic question is whether and how it can be divided into subsystems.  LQFT gives examples of such divisions, but such a division may well be part  (at least approximately) of a more complete description of quantum gravitational physics.  And if a black hole does behave as a subsystem, the problem can be phrased in terms of behavior of evolution of that subsystem.

A second point is that this presents us with what appears to be a rather clear choice, which is encapsulated in the statement of a ``black hole theorem."  If the world can be decomposed, at least to an adequate approximation, into subsystems, and quantum mechanical evolution holds, and other inconsistencies are to be avoided, that implies certain behavior of the dynamical evolution of those subsystems, which appears not to respect the standard locality constraints of LQFT.

Such a discussion, exhibiting a conflict between some very general assumptions, thus focusses attention on the key question of which of these must be modified for a consistent physical description.  Given the conflict -- and the likelihood of the problem's fundamental role -- the most important part of this ``theorem" is probably, like with the Coleman-Mandula theorem\cite{CoMa}, the loophole through which it is evaded.\footnote{The expected evasion   is one reason for the quotation marks, and another is the relative triviality of the proof, once the assumptions are explained.  To avoid clutter,  quotation marks will be dropped in most of the following, but may be mentally inserted by the reader.}  The loophole in the Coleman-Mandula theorem connects to the existence of supersymmetry, and the loophole here is expected to be associated with  deeper structure and principles of quantum gravity.  

In outline, the next section will state and explain the basic assumptions, using LQFT for illustrative purposes; the theorem follows simply once these are stated.  The next section then discusses the way in which various proposals modify these assumptions. Discussion is given both for older scenarios, and particularly for the newer replica wormhole proposal.  Challenges with these proposals suggest seeking a more ``minimal" resolution of the conflict.

Section four reviews and expands on the important question of definition of subsystems in quantum gravity.  It is quite plausible that the locality property of LQFT must be modified.   An important first question is whether there is a more ``coarse-grained" notion of locality, associated to definition of quantum subsystems of a gravitational system.  In other quantum systems, such a definition of subsystems is part of the basic structure, and it is difficult to describe much of what we do in quantum mechanical theories {\it without} some such definition.  In quantum gravity the question of their definition appears to likewise be an important one of basic structure.  Assumption of such a definition is part of the statement of the general theorem, but this is clearly a question to be understood more closely. 

If a black hole can be described as a subsystem, at least to a good enough approximation to avoid providing the loophole, and other apparent inconsistencies are to be avoided, then evading the conflict requires other violations of LQFT locality.  This conclusion is examined in section five, first with corresponding restatements of the theorem, and then by connecting with work, which is reviewed, on parameterizing interactions that restore  unitary evolution.  These in turn immediately suggest possible observational signatures. 

In short, this paper exhibits a general chain of reasoning that leads to a limited set of general possibilities for resolution of the problem.  This focusses attention on key questions such as definition of subsystems in gravity.  This in turn provides additional motivation and explanation for  
 the possible presence of interactions that lie outside the standard LQFT definition of locality.

\section{``Black hole theorem," (v1)}

\subsection{Statement}\label{statement}

We immediately turn to a statement of the theorem.  Suppose that 
\begin{enumerate}
\item A black hole is a subsystem;
\item Distinct black hole states have identical exterior evolution; and
\item A black hole disappears at the end of its evolution.
\end{enumerate}

Then, the evolution of the black hole  and its surroundings violates quantum mechanics; specifically, it is not unitary.

A similar statement could be made for other examples of quantum subsystems of bigger quantum systems, such as a pair of systems $A$ and $B$ with a corresponding Hilbert space
\beq\label{Hprod}
{\cal H}={\cal H}_A\otimes{\cal H}_B\ .
\eeq
However, gravity, and black holes, have a variety of more subtle features, and so it is important to first take some time to explain these assumptions in that context.

\subsection{Explaining assumptions, and proof}

\subsubsection{Subsystems}
\label{Subs}

In some of the  literature, it is assumed that the notion of a black hole being a subsystem is captured by the statement that the full Hilbert space of the black hole together with its environment can be written as a product,
\beq\label{tensprod}
\calh = \calh_{BH}\otimes\calh_{env}\ .
\eeq
However, beginning at the level of a description of black hole evolution  via the approximation of LQFT evolution on a classical background, one sees that this structure is likely problematic, due to the infinite entanglement between regions in LQFT,\footnote{This is associated with the type III property of von Neumann algebras in LQFT, as for example explained in reviews such as \cite{Haag}.  For another discussion of this,  and of related questions of subsystems in gravity, see \cite{SGalg}.  One common belief is that gravity addresses the problem of infinite entanglement and possibly modifies the type III structure, due to the relation of corresponding short-distance divergences, for example in entanglement entropy, and renormalization of Newton's constant.  However, a complete story of how this question is addressed is not yet known.}  and consequent cutoff dependence.

\begin{figure}[!hbtp] \begin{center}
\includegraphics[width=5cm]{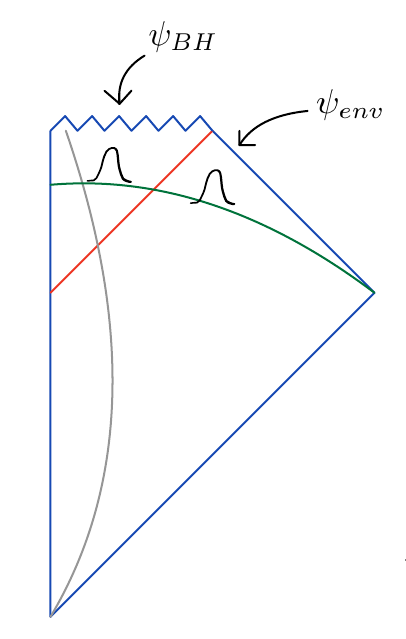}
\end{center}
\caption{A Penrose diagram for a black hole, formed from collapsing matter.  The quantum state can be specified on a slice, like that shown.  The quantum state is specified by its structure both inside the black hole, and in the exterior, as in \eqref{psitot}; one can think of independent excitations in these two regions.}
\label{FigPen}
\end{figure}

\begin{figure}[!hbtp] \begin{center}
\includegraphics[width=8cm]{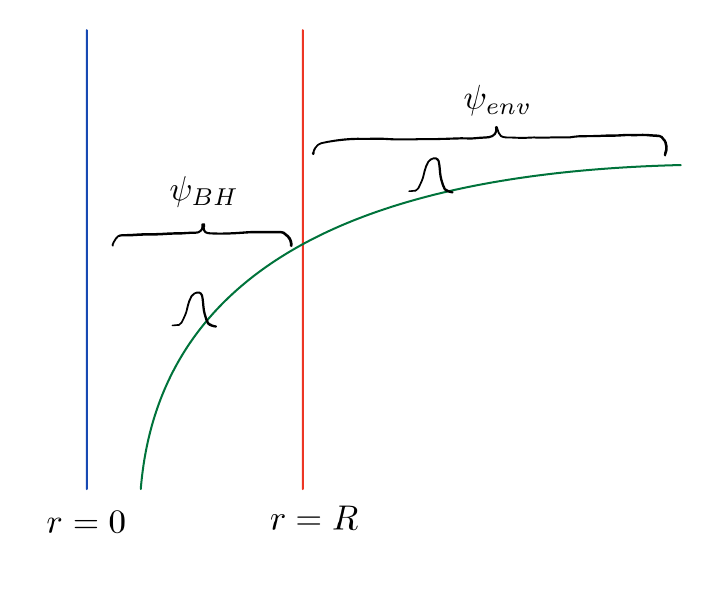}
\end{center}
\caption{An Eddington-Finkelstein diagram corresponding to the spacetime of Fig.~\ref{FigPen}.  Such a depiction has the advantage of avoiding the near-infinite conformal distortion of the Penrose diagram, late in the black hole's evolution.}
\label{FigEF}
\end{figure}

One can motivate an improved definition of subsystems from the description of the quantum state of a black hole in this approximation where it is governed by LQFT evolution on a background.  Figs.~\ref{FigPen} and \ref{FigEF} show the classical geometry of a black hole in either a Penrose or Eddington-Finkelstein representation.  The quantum state can be described on a slice like the ones shown.  The basic point is that one has much of the structure of \eqref{tensprod} in that the state can be written
\beq\label{psitot}
|\Psi\rangle = |\psi_{BH},\psi_{env}\rangle
\eeq
where we have independent labeling and description of the parts of the state inside the black hole and in the environment.  Another way of capturing this structure, used in the algebraic quantum field theory literature (see, {\it e.g.} \cite{Haag}), is to describe the commuting operator subalgebras $\cala_{BH}$ and $\cala_{env}$ that act separately on the two corresponding parts of the state.\footnote{Yet another way to describe subsystems, which leads to a tensor factorization like \eqref{tensprod}, albeit for modified states, is to use the split vacuum (see, {\it e.g.}, \cite{Haag}).}

Extending to include gravitational backreaction, we find that external attributes of the state depend on some of the properties of its internal part, like the total energy.  Generally, we might assume that there are different internal states that lead to the same total Poincar\'e charges.\footnote{This is clearly possible in the presence of global symmetries, but may also be possible more generally.}  In that case, the exterior part of the state is expected to depend on the values of these charges, since they are asymptotically observable via the gravitational field\cite{DoGi4,SGsplit}.  We will return to further discussion of the more complete situation later in the paper, but for now work using the description \eqref{psitot} or its generalization to include such charge dependence.  We emphasize that while the above discussion of LQFT provides such a structure, we take that as only an {\it example} of such a structure of a more general framework.

This is a good place to stress a point that appears likely to be important: in other quantum systems in physics, such as finite quantum systems or those of LQFT, such subsystem structure is {\it hardwired at the beginning} -- it is part of the basic underlying description of the physics.  This raises the important question of whether such subsystem structure is part of the fundamental description in quantum gravity, and if so whether it is mathematically characterized as we have described or by another mathematical 
structure\cite{SGalg}\cite{QFG}\cite{DoGi4,SGsplit} -- or whether gravity is somehow fundamentally different in this regard.

\subsubsection{Identical exterior evolution}

\begin{figure}[!hbtp] \begin{center}
\includegraphics[width=9cm]{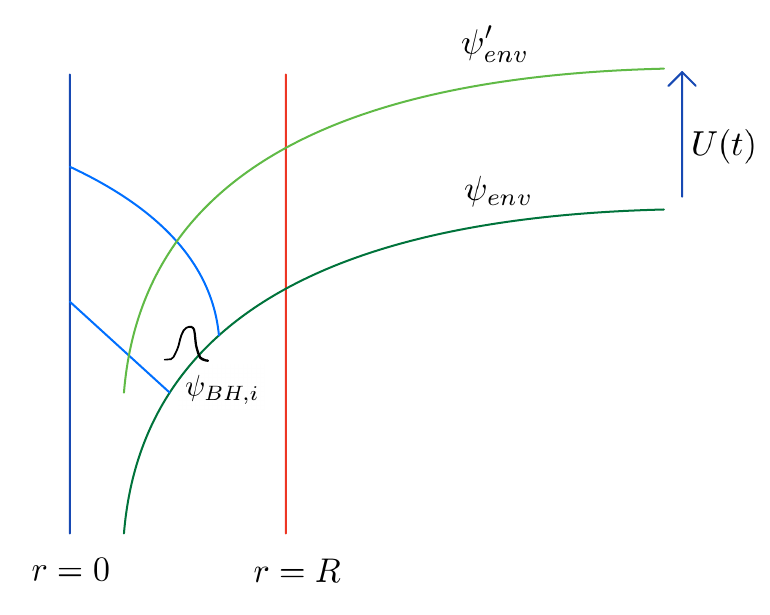}
\end{center}
\caption{Evolution of the quantum state between slices.  In a description based on LQFT, the internal excitation stays within the light trajectories (shown), and thus does not affect the external part of the state.}
\label{Figslices}
\end{figure}

The property of identical exterior evolution can also be illustrated in the LQFT approximation, but stated in the more general context.  Fig.~\ref{Figslices} shows two slices through a black hole geometry, with possibly different internal components $\psi_{BH,i}$, or excitations, labelled by $i$, of the state on the first slice.  If one considers LQFT evolution between the slices -- which can be very concretely described in terms of a hamiltonian,\footnote{See \cite{SGsch,SG2d,GiPe} for recent explicit treatment of such hamiltonians.} then the differences in the internal state do not affect the evolution of the external part of the state:  the evolution operator $U(t)$ maps 
\beq\label{indepev}
|\psi_{BH,i},\psi_{env}\rangle\ \rightarrow\ |\Psi(t)\rangle =|\psi_{BH,i}',\psi_{env}'\rangle
\eeq
where
the external $\psi'_{env}$ is independent of $i$.  The different $i$ only affect the evolution inside their forward lightcone, which is shown, and evolves to $r=0$.  While LQFT motivates the structure of the evolution \eqref{indepev}, that structure is more general.

One way to imagine characterizing this independence is by calculating a density matrix,
\beq
\rho_{env} = Tr_{BH}|\Psi(t)\rangle\langle\Psi(t)|
\eeq
which if this condition holds is expected to be independent of $i$.  However, defining such a density matrix depends on introducing some cutoff prescription.  An alternative is to consider operators $\calo_{env}$ acting on the environment, and state that
\beq
\langle\Psi(t)|{\cal O}_{env}|\Psi(t)\rangle
\eeq
is independent of the black hole state $i$ for all such operators in the observable subalgebra $\cala_{env}$.  

In the example of description in terms of  LQFT on a background, this property of identical exterior evolution follows from the locality property of LQFT, which states that operators acting to change the inside part of the state on the initial slice will always commute with  future operators acting on the environment, due to the light-cone structure of the background.  This can alternately be described in terms of the hamiltonian.  One might ask whether it remains true with dynamical geometry.  In fact, one can substantially account for such dynamics by semiclassically including the effects of the backreaction such as in \cite{CGHS}.  The description continues to exhibit the property of external independence of the state, arising from the presence of an effective horizon in the semiclassically corrected geometry (see, {\it e.g.} \cite{GiNe} for some discussion).

\subsubsection{Disappearance}

By disappearance, we mean that the black hole decays and after its decay time is not 
 part of the Hilbert space, so the evolution operator $U(t)$  maps
\beq\label{finstate}
|\psi_{BH,i},\psi_{env}\rangle\ \rightarrow\ |\psi_{env}''\rangle
\eeq
at sufficiently late times.  The identical exterior evolution property means that the later state is independent of $i$.\footnote{An alternative is that this property breaks down before whatever is left of the black hole disappears, that is we have a black hole remnant that radiates attributes of the initial state.  This possibility will be considered in the next section.}

At this point the ``proof" of the theorem is essentially trivial: evolution of the form \eqref{finstate} is many to one, and is therefore not unitary, so violates quantum mechanics.

There are various diagnostics of this.  For example, one can consider a pure state that entangles the internal states of the black hole with an auxiliary system with states $|\tilde i\rangle$, such as
\beq
\sum_i|\psi_{BH,i},\psi_{env}\rangle |\tilde i\rangle\ .
\eeq
In fact, LQFT evolution produces such a state via the Hawking process, with the entanglement being with the earlier Hawking radiation described by $|\tilde i\rangle$.  This entanglement  can be characterized by the von Neumann entropy found by tracing out the state of the black hole to give a nonzero, and ultimately large, result.  If the true evolution were  however of the form \eqref{finstate}, that would  produce a final entropy that vanishes, and this difference characterizes the magnitude of the violation of quantum unitarity.

An obvious alternative is to abandon quantum mechanics.  However, approaches to its generalization have typically led to disaster\cite{BPS,Polphone}, and specifically approaches that systematically generalize field theory evolution to incorporate the kind of nonunitarity we have described appear to lead to massive violation of energy conservation\cite{BPS}.

So, to evade the theorem, and restore the primacy of quantum mechanics, one apparently needs to violate one or more of the assumptions listed in \ref{statement}.  Like with the Coleman-Mandula theorem, the most important and interesting aspect is likely the loophole in the reasoning, which we expect may guide our understanding of the new principles of quantum gravity.  Specifically, we can try to parameterize or otherwise characterize the needed violations, and seek clues from these about the deeper structure of the physics.

\section{Proposals for evasion}

There have been various proposed escapes from the reasoning surrounding the problem.  A good test question for better understanding and characterizing the nature of such scenarios is to ask how specifically a scenario differs from the LQFT evolution which produces the Hawking state --  which has been studied in increasing detail\footnote{For recent work to describe the evolving Hawking state, see \cite{SGsch,SG2d,GiPe}.} -- through modification of the Hilbert space, hamiltonian, or other aspect of the structure.  One approach to characterizing this difference is to understand what a scenario says about modification of the above assumptions.

\subsection{Some older scenarios}

One longstanding proposal -- that is still considered (see {\it e.g.} \cite{Bianchi:2018mml}) -- is that a black hole leaves behind a long lived remnant which contains the information about its earlier quantum state.  This would violate assumption 3), disappearance.  General arguments based on energy conservation constraints on information transfer rates\cite{CaWi,Pres} indicate that such a remnant must be extremely long-lived, with lifetime $\propto (M/m_{Pl})^4$ in terms of the initial mass $M$ of the black hole.   This scenario introduces other extremely serious problems\cite{WABHIP,Susstrouble} due to the unbounded number of quantum states of such remnants, such as unbounded production rates in general physical processes.

A different general scenario with various realizations is that at some point earlier in its evolution, long before it reaches the Planck mass, a black hole transitions to a new kind of object which can be called a ``massive remnant" \cite{BHMR}, and might for example be thought of as an exotic starlike object.  There are multiple proposals for such objects at varying levels of detail:  gravastars\cite{MaMo}, fuzzballs\cite{fuzzrev}, firewalls\cite{AMPS}, Planck stars\cite{RoVi}, {\it etc.}  Such a scenario represents a rather drastic departure from the usual description of black holes, and typically requires a collection of other assumptions, including some form of nonlocality.
This kind of scenario would modify assumption 1), subsystems, as well as assumption 2), identical exterior evolution.

Another proposal is that of ER=EPR\cite{MaSu}.  This suggests a different kind of subsystem identification.  However, an open question is how to systematically account for this within quantum mechanics.

There are various other scenarios, for example that of `t Hooft\cite{thooft}.  Describing these in terms of how they modify LQFT evolution, and depart from our assumptions, is left as an exercise for the reader.

\subsection{Replica wormholes}

\begin{figure}[!hbtp] \begin{center}
\includegraphics[width=9cm]{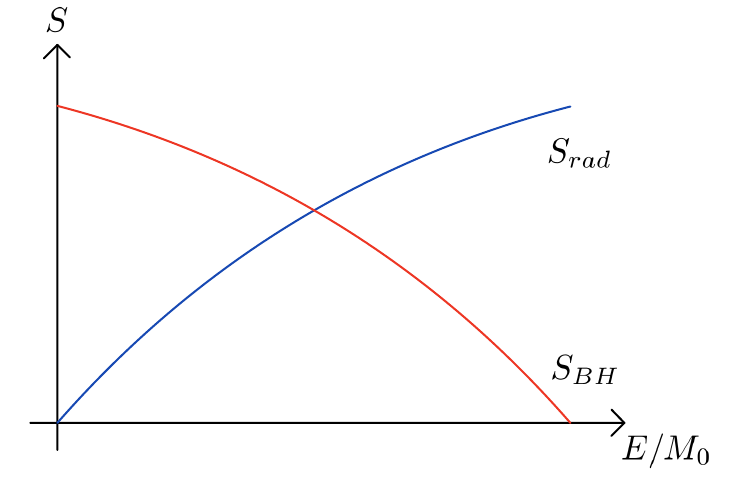}
\end{center}
\caption{A sketch of curves representing the growing entanglement entropy of outgoing Hawking radiation with the internal states of a black hole, and the falling Bekenstein-Hawking entropy.  If black holes behave like other quantum subsystems, it has been argued the true von Neumann entropy of outgoing radiation should approximately follow the lower of the two curves.}
\label{Figent}
\end{figure}

A more recent proposal is the replica wormhole proposal\cite{AEMM,AMMZ,PSSY,AHMST,AHMSTrev}.  We have long understood the curves shown in Fig.~\ref{Figent}, representing the growth of entanglement of the black hole state with outgoing Hawking radiation, as derived from LQFT, and the falling Bekenstein-Hawking entropy.  A general argument, apparently first clearly stated in print by Page\cite{Pageav,Pageinfo}, says that if a black hole behaves like a quantum system, disappears, and evolution is unitary, then the actual von Neumann entropy of the outgoing radiation is expected to closely follow the composite curve that is the lower of the two.

In short, the replica wormhole story proposes a prescription to choose the lower of the two known curves, which is motivated by the observation that they can correspond to different saddlepoints in euclidean geometries, generalizing standard replica calculations.

However, von Neumann entropy is just one diagnostic, and reproducing known curves is not convincing without a more complete story:  we would specifically like to relate this proposal to the standard quantum-mechanical framework of states and amplitudes.  Another possibility -- briefly examined in \cite{GiTu} -- is that this story represents a {\it modification} of standard quantum-mechanical rules, for example by introducing new rules that generalize taking traces of products of density matrices or identification of subsystems; in that case, we would like to systematically understand the more complete logical structure.

\begin{figure}[!hbtp] \begin{center}
\includegraphics[width=4cm]{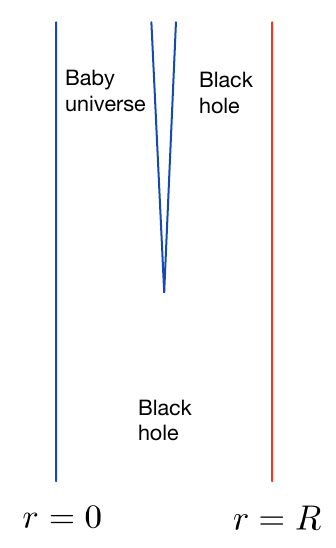}
\end{center}
\caption{Depiction of a process in which a baby universe is emitted from the interior of a black hole.}
\label{Figbu}
\end{figure} 

If we focus on  more standard quantum-mechanical rules, a proposed picture is the connection to baby universes and spacetime wormholes\cite{GiStinst,Cole,GiStinc,GiSt3Q} first explored by Marolf and Maxfield\cite{MaMa,MaMarev}.  A rough picture of what could be happening is sketched in Fig.~\ref{Figbu}: a separate baby universe branches off from the black hole.  This kind of process certainly suggests a modification of the subsystem structure, and moreover suggests that information can go off into -- or return from -- a state of multiple baby universes, emitted by black holes\cite{LRT,Hawkworm}.

However, at least working in an approximation, a more careful analysis\cite{Cole,GiStinc} found that this is not the case.  One can summarize the interaction of a baby universe with our universe by a correction to the hamiltonian
\beq\label{deltah}
\Delta H = (a_{BU,A}+a_{BU,A}^\dagger) \, {\cal O}_A\ 
\eeq
where $a_{BU,A}$ and $a_{BU,A}^\dagger$ annihilate or create baby universes of type $A$, and $\calo_A$ represents the effect on the state of our universe of their absorption or emission.  Then, we find states of the baby universes, called alpha vacua, so that
\beq
(a_{BU,A}+a_{BU,A}^\dagger)|\alpha_A\rangle=\alpha_A|\alpha_A\rangle\ .
\eeq
In such a state, the correction \eqref{deltah} simply introduces new coupling constants to the operators $\calo_A$; the presence of a nontrivial superposition of states means that there is a probability distribution for these couplings.  If we focus on an alpha vacuum, and if the operators $\calo_A$ are localized inside the black hole, they do not appear to change the conclusions of the theorem -- specifically they do not affect the external evolution.  

On the other hand, this story does suggest a possible way to evade the theorem, if there are important contributions from operators $\calo_A$ that are nonlocal on scales $\gtrsim R$, corresponding to large wormholes.  These can violate assumption 2), identical exterior evolution.  We will return to this type of scenario later.

Another possibility is that the preceding approximation, based on neglecting baby universe interactions, fails in an important way.  Or, possibly there is some other basic explanation of the amplitudes underlying the replica wormhole prescription for reproducing the entropy.  If so, it is important to systematically understand what it is.

\subsection{A more minimal scenario?}

The scenarios described above typically involve considerable extra structure, and a range of further nontrivial assumptions which may be hard to satisfy.  This suggests that we return to the assumptions behind the theorem, and examine them closely to see whether there might be a less drastic and elaborate way to escape their conclusions, with a minimum of new assumptions.

\section{The question of subsystems in quantum gravity}

The theorem clearly applies with a subsystem structure like in the finite case, \eqref{Hprod}, or as in LQFT.  However, 
defining subsystems is more subtle in quantum gravity than in typical finite systems or LQFT, and this suggests more careful examination of the subsystem assumption and the possibility of different behavior.  
As noted above, a definition of subystems is part of the defining structure of most complex quantum systems, and so we might expect that to also be the case for quantum gravity.  We can begin to understand the subtleties after first reviewing definitions in simpler cases.

In finite systems, or locally finite systems such as lattice field theories, subsystems can be mathematically defined in terms of a tensor factorization of the Hilbert space, as in \eqref{tensprod}.  Physically, this implements the condition that  an observation or measurement on one of the subsystems, say $B$, is independent of the state in the other subsystem, say $A$.  

In LQFT, one na\"\i vely expects such a factorization corresponding to different regions of spacetime, but the story is more subtle due to the type III property, and corresponding infinite entanglement.  For this reason, one instead proceeds by associating subsystems to commuting subalgebras of the algebra of observables, corresponding to observables localized in different regions.  This definition is developed in the algebraic quantum field theory literature, see {\it e.g.} \cite{Haag}.  (See also \cite{SGalg} for further discussion.)

A simple example of the role of this algebraic structure  can be given for a scalar field $\phi(x)$.  Consider a state given by
\beq\label{JstateD}
|J\rangle= e^{-i\int J(x)\phi(x)} |0\rangle = {\cal O}_J|0\rangle
\eeq
where the function $J(x)$ has compact support restricted to some neighborhood $U$.  Then, observables, such as the field $\phi(x)$, are independent of $J$ if they are spacelike separated from $U$.  A concrete example is the statement
\beq\label{Jstate}
\langle J|\phi(x)|J\rangle=\langle0|\phi(x)|0\rangle
\eeq
for $x$ spacelike to $U$.  This is a starting point for describing locality of LQFT. Hawking evolution respects this locality, and that is what forbids information escape in the Hawking process.

Extending to gravity, the problem is that field observables such as $\phi(x)$ are not gauge invariant and thus not physical; diffeomorphisms act nontrivially on them.  This can also be understood by the failure of $\phi(x)$ to commute with the constraints, which in an effective description of quantum general relativity take the form
\beq
C_\mu(x)=\frac{1}{8\pi G} G_{0\mu}(x)-T_{0\mu}(x)\ ,
\eeq
and generate  diffeomorphisms.
Likewise, the state $|J\rangle$ is not annihilated by the constraints.  

The physical reason for this is that a particle is inseparable from its gravitational field.  The solution is to solve for operators commuting with the constraints, which will include a ``gravitational dressing" describing this field.

This has been done at the perturbative level in \cite{DoGi1,DoGi4} in a Minkowski background, and \cite{GiKi} for an AdS background (see also \cite{CGPR}).\footnote{Earlier related work includes \cite{Heem} and  \cite{KaLigrav}.   The former exhibited nontrivial commutators arising from the constraints but did not describe the dressed operators; the latter was focused on finding bulk operators that {\it commute}, and did not exhibit  the bulk dressing and the noncommutative behavior described below.}  The metric operator is written as a perturbation $h$ about a background metric,
\beq
\tilde g_{\mu\nu} = g_{\mu\nu} + \kappa h_{\mu\nu}\ ,
\eeq
with $\kappa^2=32\pi G$.  Then, for example, the state \eqref{JstateD} is promoted to a gravitationally-dressed state involving the metric perturbation
\beq
|J\rangle= {\cal O}_J|0\rangle\rightarrow |\widehat J\rangle =\hat{\cal O}_J[\phi,h]\,|0\rangle\ ,
\eeq
with square brackets  here indicating functional dependence (not the commutator).

The constraints are solved, to leading order in $\kappa$, by expressions of the form
\beq
|\widehat J\rangle \simeq e^{i\int V^\mu[h]\,T_{0\mu}} |J\rangle
\eeq
or
\beq\label{dressop}
\hat {\cal O}_J \simeq e^{i\int V^\mu[h]\,T_{0\mu}}\, {\cal O}_J\, e^{-i\int V^\mu[h]\,T_{0\mu}}\ .
\eeq
Here the integrals are spatial volume integrals and
the  $V^\mu[h]$ are functionals of the metric perturbation, which can be explicitly written down.  In fact, there are many such functionals that can be chosen, corresponding to the choice of different instantaneous gravitational fields of a matter distribution; these differ by pure gravitational radiation.  An example in a flat background is the line integral expression \cite{DoGi1,QGQF}
\beq
 V_\mu^\Gamma(x)= \frac{\kappa}{2} \int_x^\infty dx^{\prime\nu} \left\{ h_{\mu\nu}(x') + \int_{x'}^\infty dx^{\prime\prime\lambda}\left[\partial_\mu h_{\nu\lambda}(x'') - \partial_\nu h_{\mu\lambda}(x'')\right]\right\}
\eeq
where $\Gamma$ is a curve connecting $x$ to infinity.
The composite operator \eqref{dressop} can be though of as creating the underlying matter excitation, together with such a gravitational field.

Now, however, the operators corresponding to spacelike separated sources $J_1$, $J_2$ generically no longer commute\cite{SGalg,DoGi1},
\beq
[\hat{\cal O}_{J_1}\,,\hat{\cal O}_{J_2}]\neq0\ ,
\eeq
due to terms from the dressings, which must extend to infinity\cite{DoGi2}.  Likewise, the metric perturbation operator (which might be taken to also be a dressed operator) at spacelike separations to a source   now depends on the presence of the source,
\beq
\langle\widehat J|\hat h_{\mu\nu}(x)|\widehat J\rangle \neq\langle 0|\hat h_{\mu\nu}(x)|0\rangle\ ,
\eeq
also due to the dressing.  So, this raises the question, in what sense does information localize in quantum gravity?  Or, what is the definition of a subsystem in quantum gravity?  As suggested above, this looks like a key {\it structural} question for quantum gravity.

While the complete nonperturbative story has not been fully described, there are some results.  First, the leading perturbative dressing can be chosen, using the freedom of choosing $V^\mu[h]$ described above, so that $n$-point functions of the metric perturbation spacelike to a source,
\beq
\langle\widehat J|h_{\mu_1\nu_1}(x_1)\cdots h_{\mu_n\nu_n}(x_n)|\widehat J\rangle
\eeq
only depend on the total Poincar\'e charges (or their moments)\cite{SGsplit} of the source $|\hat J\rangle$.\footnote{This also suggests that \eqref{psitot} be generalized, as noted in \ref{Subs}, to include the eigenvalues of a maximal commuting set of Poincar\'e generators, which label the state and are observable via the asymptotic metric.}  This, in particular, strongly suggests\cite{DoGi3,DoGi4,SGsplit} that soft hair\cite{Hawk-Info,HPS1,HPS2,astrosrev,astrorevisit,HHPS} doesn't play an important role in encoding information, since the soft charges have little {\it necessary} correlation with the state in an interior region.

On the other hand, there are asymptotic observables whose values depend on aspects of the dressed state $|\hat J\rangle$.  To understand this, note that the translation generators can be written in the form
\beq
P_\mu= P_\mu^{ADM}[h(\infty)] + \int dV C_\mu
\eeq
where $P_\mu^{ADM}$ are the ADM expressions, and only depend on the asymptotic metric.  Then, if one has set $C_\mu=0$ by solving the constraints, the translation generators are purely asymptotic (or boundary, in the case of AdS) expressions, and have nontrivial values for dressed states. 

The leading candidate for an explanation for holography has been suggested to follow from this, by Marolf\cite{MaroUH,Marothought, MaroholoNS}.  If one considers the state $|\hat J\rangle$ corresponding to a matter field excitation restricted to a neighborhood $U$ at time $t=0$, or the corresponding operator $\hat\calo_J$, then in AdS it is argued this state or operator may be related to a state or operator on the boundary of AdS to the causal future of the neighborhood $U$ through the map given by propagation of the excitations.  If this is the case, it is then argued that the hamiltonian, which is now a pure boundary term, may be used to relate the state or operator on the boundary to an earlier one at $t=0$, and that this constructs the ``holographic map" between bulk and boundary states or operators.

In fact, given the preceding discussion, there is a simpler candidate for relating localized states or operators to asymptotic observables, given in \cite{DoGi3}:  one can act on the state $|\hat J\rangle$ by a translation generator $P_i^{ADM}$ which is purely an asymptotic operator, to translate the excitation out to the asymptotic region, and then detect it via an asymptotic observable.  Explicitly, one can for example consider the expression
\beq
\langle\widehat J|{\cal O}(x)\, e^{i P_i^{ADM} a^i} |\widehat J\rangle \ ,
\eeq
where $\calo(x)$ is an observable in the asymptotic region far from $U$, for example at the boundary of AdS.  This expression thus involves expectation values only of asymptotic operators, yet by the preceding argument is sensitive to the structure of the state, since for large enough $a^i$, $P_i^{ADM}$ translates the state to the asymptotic region, where $\calo(x)$ can detect it.

These arguments raise the question whether information is delocalized in gravity; certainly the description of its localization is more subtle.  However, there are some important comments to bear in mind.  First, one needs to have a solution of the constraints $C_\mu=0$, or their generalization in the more complete theory, to make these arguments.  Moreover, this apparently needs to be a solution of the constraints at a {\it nonperturbative} level.  One way to see this is to consider states that correspond to the operators
\beq
e^{-i\int J(x)\phi_i(x)}
\eeq
for two different scalar fields $\phi_1$, $\phi_2$, and moreover consider these acting to create excitations inside a black hole.  Clearly to apply the preceding arguments about translating and observing, or propagating to infinity, translating, and observing, one needs non-perturbative control over the solution to the constraints.  In fact, even without considering states inside black holes, the preceding arguments require acting to translate the state or operator over distances that are very long as compared to the Planck length, or even infinite, and this also appears to indicate the need for a construction going beyond leading perturbative level. Ref.~\cite{LPRS} has argued that  such constructions extend to imply {\it perturbative} availability of information, but in simple examples the resulting effect is exponentially small in the magnitude of the translation.

In short, given these limitations, a question remains regarding in what sense information is effectively delocalized in gravity.  And, specifically, we would like to understand if this delocalization, say described in perturbative gravity, can be sufficient to effectively transfer  information outside a black hole,\footnote{One approach to a more precise characterization of such information transfer is via entanglement transfer\cite{GiSh1,Sussxfer}.}
and evade the theorem, {\it e.g.} by sufficient violation of assumption 2), or whether new non-perturbative structure is needed.  In fact, the relation to the constraints suggests a certain circularity, if one wants to use the preceding discussion of holography to argue for a resolution of the black hole conflict: solving the constraints is tantamount to describing the unitary evolution, and so one needs to understand this evolution to construct the holographic map\cite{HoUn}.

\section{Restatements, and implications}

We have found that the assumption of unitary quantum evolution, which we can think of as assumption zero, and assumptions 1)-3) of 
\ref{statement}, are inconsistent.  There are different ways to organize the statement of the conclusion.  A first is that assumption 0), quantum evolution, together with assumption 3), black hole disappearance, imply that assumptions 1) or 2) or both fail.  This version two of the theorem 
can be regarded as a ``nonlocality theorem," stating the departure of the physics from the LQFT paradigm.

However, it remains plausible that there is, within quantum gravity, a description of subsystems, at least to a sufficiently good approximation for purposes of describing evolving black holes.  If one also makes this assumption, then the combined assumptions 0), 1), and 3) imply that assumption 2) must fail, that is distinct black hole states {\it don't} have identical exterior evolution.  This can be regarded as a third version of the theorem.

\subsection{Parameterizing black hole quantum interactions, and possible observational windows}

This last version of the theorem implies that there {\it must} be interactions of the black hole with the environment that depend on the quantum state of the black hole.  This conclusion has been investigated in a series of papers\cite{SGmodels,BHQIUE,GiSh1,NVNL,NVUEFT,GiSh2,NVNLT,SGObs,NVU}.  Such black hole state-dependent interactions violate locality of the semiclassical spacetime, but it is not clear they are problematic from a more basic quantum viewpoint.  Indeed, if there is an error in the semiclassical description of the spacetime, as compared to a more fundamental quantum description, these interactions may arise in the description of this error when one uses the semiclassical description.  Alternately, they might arise in parameterizing the effects of corrections to an approximate subsystem decomposition.  In any case it is important to understand what kind and size of interactions is needed to restore unitary evolution.

One can explore such interactions in a principled parameterization of ignorance.  Such parameterizations can have remarkable power; an example is the derivation of the field theory of the Standard Model via such a parameterization, after assuming the relevant symmetries.  More discussion appears in \cite{BHQU} and the other references, but a brief summary will be given here.  We work about the approximation in which the black hole and environment are subsystems, and parameterize the necessary interactions.

If we consider a quantum state of black hole plus environment, $|\Psi\rangle \sim |\psi_{BH,i},\psi_{env}\rangle$, the LQFT evolution will give identical exterior evolution, so must receive a correction, so the hamiltonian is of the form
\beq
H=H_{LQFT} + \Delta H\ .
\eeq
 The simplest form of an interaction term that transfers information from the black hole state to the environment is a bilinear of operators acting on each of these subsystems.   If we assume there are finitely many black hole states, indexed by $i,j,...$, then the general such operator is of the form
 \beq
 \Delta H = \sum_A \lambda^A_{ij} {\cal O}^{env}_A
 \eeq
 where the $\lambda^A$ give a basis for operators on the black hole subsystem and where $\calo_A^{env}$ act on the environment.  If we assume that the state of the environment is to a good approximation described by LQFT, as expected for the exterior of a large black hole, then the latter can be well-approximated in terms of operators of LQFT.

One can study the effect of $\Delta H$ for general such operators\cite{NVUEFT,GiSh2}, but a specific form for them can be strongly motivated.  

First, a natural scale on which these operators are expected to act is that of the event horizon size $R$.  From the perspective of the semiclassical description these operators must act outside the event horizon,\footnote{Here we make contact with the earlier baby universe discussion.} but requiring them to, say, only act in a vicinity of the event horizon of Planck thickness would require finely tuning them, and moreover produces an unnatural consequence, a firewall\cite{SGTrieste,Brau}\cite{AMPS}.   An additional motivation is that there are multiple arguments\cite{SGBoltz,DLP}\cite{SGsch,SG2d,GiPe} that Hawking radiation is produced in a ``quantum atmosphere" of thickness $\calo(R)$; this suggests that this is the domain relevant also to resolution of the problems associated with Hawking radiation.   Likewise, we expect typical transitions to be induced between states with energy difference $\sim 1/R$.

Second, one reasonably hypothesizes that $\Delta H$ couples universally to {\it all} fields.  This helps preserve important aspects of the beautiful story of black hole thermodynamics.  Specifically, democratic radiation of all species is a hallmark of thermal systems, and while such couplings can increase the total radiation rate, this can possibly be reconciled with the Bekenstein-Hawking formula for the entropy if interpreted as stemming from an increase of the effective area in the Stefan-Boltzmann law\cite{SGBoltz}.  Universal couplings also address\cite{NVNL,NVUEFT,NVNLT} Gedanken experiments of black hole mining\cite{UnWamine,LaMa,FrFu,Frol}, in which through introduction of additional structure, such as a cosmic string, a black hole's decay rate increases; there must be commensurate increase in the information transfer from the black hole, to avoid reproducing the basic conflict.  A final motivation is the observed universality of gravitational phenomena.

These assumptions can be satisfied with an expression
\beq\label{DHT}
\Delta H = \sum_A \lambda^A \int dV G_A^{\mu\nu}(x)\,T_{\mu\nu}(x)
\eeq
where $T_{\mu\nu}$ is the stress tensor (including perturbative gravitons), where the integral includes the exterior of the black hole, and where $G^{\mu\nu}_A(x)$ can be thought of as ``form factors," and have support and spatial variation on scales $r\sim R$.  These are part of the parameterization of ignorance, and should be ultimately derived from a more fundamental description of the states and dynamics of the theory.  
The hamiltonian correction \eqref{DHT} can be reorganized in the form
\beq
\Delta H = \int dV\, H^{\mu\nu}(x) \,T_{\mu\nu}(x)\ ,
\eeq
where 
\beq
H^{\mu\nu}(x)=\sum_A\lambda^A G_A^{\mu\nu}(x)
\eeq
behaves like a perturbation of the metric, but is an operator acting on the black hole state.

A final, critical, constraint on the couplings $G_A^{\mu\nu}(x)$ is that they need to transfer information (or entanglement) from the black hole state at a rate $\calo(1/R)$.  This is in order to compensate for the buildup of entanglement of the Hawking process, governed by $H_{LQFT}$, at a rate of this size. 

If we consider the expectation value in a typical black hole state,
\beq
\langle \psi_{BH},t|H^{\mu\nu}(x)|\psi_{BH},t\rangle\ ,
\eeq
the preceding assumptions imply that it varies on spatial and temporal scales $\sim R$.  If it also is typically of size $\calo(1)$, then that will clearly be sufficient to produce an $\calo(1)$ modification of the Hawking state and accomplish information transfer at a rate of size $1/R$.  This possibility can be called a strong or coherent scenario, since the metric perturbations then behave like a classical $\calo(1)$ perturbation to the metric.  Since generically these are both space and time dependent on scales $\sim R$, they are expected to lead to time-dependent distortions of electromagnetic images of black holes\cite{SGObs,SGAstro,GiPs} which could be visible to observations such as with the Event Horizon Telescope (EHT).  Recent observations of the image of the central object in M87\cite{EHT}, with reasonable agreement with the predictions of classical general relativity, are thus already beginning to place bounds on this kind of scenario.

While the strong/coherent scenario, if realized in nature, could ultimately lead to such spectacularly visible effects, an important question is what size of the effective metric perturbations $H_{\mu\nu}$ is {\it necessary} to transfer information at the required rate.  This question makes contact with a general question in information theory: given two subsystems $A$ and $B$, with a coupling between them, and with assumptions such as sufficiently rapid thermalization of the individual systems, how rapidly is information, or entanglement, transferred from one system to another?  Ref.~\cite{NVU} made a conjecture about the rate, for weak couplings, which was further elaborated on and checked in toy models in \cite{GiRo}.  

A simple estimate of the information transfer rate is the following.  First, if a subsystem $A$ is transferring energy to $B$ through a coupling, one can approximate the information transfer rate as the rate at which transitions occur between states of $A$, exciting states of $B$.  A simple analogy is decay of an atom, where information about the excited state of the atom is transferred to the outgoing photons through the decay.  Decay rates can be well approximated by Fermi's golden rule, leading to the approximate expression for the information transfer rate
\beq\label{inforate}
\frac{dI}{dt}\sim \frac{dP}{dt}=2\pi\rho(E_f) |\Delta H|^2
\eeq
where $\rho(E_f)$ is the density of final states of the combined subsystems, and $|\Delta H|$ is the typical size of the matrix elements of the perturbation hamiltonian.

In the case where subsystem $A$ is a black hole, the density of final states includes a factor $\exp\{S_{bh}\}$, where $S_{bh}$ is a black hole entropy expected to take on a large size, {\it e.g.} comparable to the Bekenstein-Hawking entropy, $S_{BH}=A/4G$.  This means that the rate \eqref{inforate} can be of order $1/R$ if the typical size of matrix elements is\cite{NVU}
\beq
|\Delta H|\sim e^{-S_{bh}/2}\ .
\eeq
Correspondingly, the  size of the metric perturbation in a typical state is
\beq\label{incmet}
\langle H_{\mu\nu}(x)\rangle ={\cal O}(e^{-S_{bh}/2})\ .
\eeq

To estimate the effect of such perturbations on other propagating fields -- electromagnetic or gravitational -- one can modify \eqref{inforate} by, instead of considering transitions from the vacuum of the environment system to an excited state, considering transitions with a nontrivial incoming state of the field.  It then takes the form
\beq\label{newscatt}
\frac{dP}{dt}=2\pi\rho(E_f) \Bigl|\int dV \langle i| H^{\mu\nu}|\psi\rangle\langle \beta| T_{\mu\nu}|\alpha\rangle\Bigr|^2\ ,
\eeq
where $|\alpha\rangle$ and $|\beta\rangle$ denote the incoming and outgoing states of the environment.  The scalings of $\rho(E_f)$ and $\langle H_{\mu\nu}\rangle$ are generically expected to remain as above, again resulting in an $\calo(1/R)$ rate.  
The preceding assumption that $R$ sets the spatial scale of variation implies that the typical momentum transfer in such a transition is also of size $\sim 1/R$.  If one considers photons observed by EHT, with wavelengths $\calo(mm)$, and a black hole size for M87 of $R\sim 2\times 10^{10}\, km$, this looks utterly negligible.  However, if one instead considers gravitational radiation from mergers, for example seen by LIGO/VIRGO, part of the signal is at wavelengths $\sim R$.  Eq.~\eqref{newscatt} implies modifications to the absorption and reflection of the signal at these wavelengths, suggesting possible sensitivity of gravitational wave observation to these effects\cite{SGsearch}\cite{BHQU}, which are under further exploration.

\subsection{Summary}

In short, the assumption that black hole evolution fits within the rules of quantum mechanics is a powerful one.  When combined with the assumption that black holes disappear at the end of their evolution, to avoid apparently disastrous consequences of microscopic remnant scenarios\cite{WABHIP,Susstrouble}, this appears to {\it imply} violation of the locality of local quantum field theory.  There are various attempts to avoid this problem which involve significant and typically problematic extra structure; an important question is whether one can do so with minimal new assumptions and without dramatic new structure.
The question of how to characterize subsystems -- and thus give at least a coarse-grained, possibly approximate characterization of locality -- in quantum gravity is an important one, that appears relevant to the basic structure of the theory of quantum gravity.  If a black hole can be described as such a subsystem, to a good enough approximation, then there must be interactions between the black hole and its environment that go beyond a standard local field theory description.  Since these need to extend outside the horizon, and moreover the natural scale associated with the strong gravity region is $\sim R$, this indicates the possibility that they affect propagating radiation in the black hole vicinity, resulting in effects on observable signals.

\vskip.3in
\noindent{\bf Acknowledgements} 
 
This material is based upon work supported in part by the U.S. Department of Energy, Office of Science, under Award Number {DE-SC}0011702, and by Heising-Simons Foundation grant \#2021-2819.

\mciteSetMidEndSepPunct{}{\ifmciteBstWouldAddEndPunct.\else\fi}{\relax}
\bibliographystyle{utphys}
\bibliography{bhthm}{}

\end{document}